\documentclass[preprint,12pt]{elsarticle}
\usepackage{graphicx}
\usepackage{amssymb}
\usepackage{amsmath}

\journal{Nuclear Physics A}

\begin{document}

\begin{frontmatter}

\title{Antinucleon-nucleus interaction near threshold from the Paris 
$\bar NN$ potential}

\author[a]{E.~Friedman\corref{cor1}}
\author[a]{A.~Gal}
\author[b]{B. Loiseau}
\author[c]{S. Wycech}
\cortext[cor1]{Corresponding author: E. Friedman, elifried@cc.huji.ac.il} 
\address[a]{Racah Institute of Physics, The Hebrew University, 91904 
Jerusalem, Israel} 
\address[b]{Sorbonne Universit\'{e}s, Pierre \& Marie Curie et Paris Diderot, 
IN2P3-CNRS, Laboratoire de Physique Nucl\'{e}aire et de Haute \'Energies, 
Groupe Ph\'enom\'enologie, 
\newline
4 place Jussieu, 75252 Paris, France.}
\address[c]{National Centre for Nuclear Studies, Warsaw, Poland}

\begin{abstract}
A general algorithm for handling the energy dependence of hadron-nucleon 
amplitudes in the nuclear medium, consistently with their density dependence, 
has been recently applied to antikaons, eta mesons and pions interacting 
with nuclei. Here we apply this approach to antiprotons below threshold, 
analyzing experimental results for antiprotonic atoms across the periodic 
table. It is also applied to antiproton and antineutron interactions with 
nuclei up to 400~MeV/c, comparing with elastic scattering and annihilation 
cross sections. The underlying $\bar pN$ scattering amplitudes are derived 
from the Paris  $\bar NN$ potential, including in-medium modifications.
Emphasis is placed on the role of the $P$-wave amplitudes with respect 
to the repulsive $S$-wave amplitudes. 

\end{abstract} 


\end{frontmatter} 

\section{Introduction} 
\label{sec:intro} 
The connection between hadron-nucleus empirical potentials near threshold 
and the underlying hadron-nucleon interactions has been studied for years 
by analyses of strong-interaction effects in exotic atoms and in studies 
of elastic scattering of hadrons by nuclei \cite{BFG97,FGa07}. 
It was recognised in the early 1970s that the $\bar K$-nucleus interaction 
near threshold was determined by the $\bar K$-nucleon scattering amplitude 
at subthreshold energies \cite{Wyc71,BTo72,Roo75}. Recently an algorithm 
was devised to account for the subthreshold energy dependence of the 
meson-nucleon amplitude in evaluating the meson-nucleus strong-interaction 
potential in kaonic atoms and for strongly-bound $\bar K$ and $\eta$ 
mesons in nuclei \cite{CFG11,CFG11a,FGa12,GMa12,FGa13,FGM13,CFG14}. 
Pionic atoms and elastic scattering of 22 MeV $\pi^\pm$ by nuclei 
have also been studied using this approach \cite{FGa14}. 

In the present work we apply this approach to the interaction of antiprotons 
with nuclei near threshold. As in previous works, we are not interested in 
any single nuclear species but rather in global behavior. Therefore we handle 
only large data sets within `global' comparisons between calculation and 
experiment, as was done for kaonic atoms \cite{FGa13} and for pionic atoms 
and pion scattering \cite{FGa14}. Nevertheless, in order to assess the 
validity of the model, some annihilation cross sections are also considered. 
Due to the much stronger absorption of antiprotons in nuclei compared to 
pions, and even to antikaons, it is inevitable that ambiguities may exist 
in some of the conclusions. 

An extensive data-base for strong interaction effects in antiprotonic 
atoms is available from the PS209 collaboration at CERN \cite{TJC01}. 
Results for elastic scattering of 48 MeV antiprotons on C, Ca and Pb nuclei 
are available from the pioneering experiments of the 1980s \cite{JLG86}. 
For the free-space ${\bar p}N$ interaction near threshold we used the 2009 
version of the ${\bar N}N$ Paris potential~\cite{ELL09}. This potential 
consists of a long-range one-pion and correlated two-pion exchange terms, 
plus a short-range phenomenological term that includes an absorptive 
component representing ${\bar p}N$ annihilation. The potential parameters 
are fitted to some 4300 scattering data plus scattering lengths and 
scattering volumes extracted from antiprotonic hydrogen levels. This 
qualifies the 2009 version of the ${\bar N}N$ Paris potential as a realistic 
potential. Other realistic, and `microscopic' as well, ${\bar N}N$ potential 
models that have become available recently could, in principle, be used 
in ${\bar p}$-nucleus calculations near threshold. These include (i) the 
Zhou--Timmermans model \cite{ZTi12} which is also based on a long-range 
one-pion and correlated two-pion exchange terms, but uses a boundary 
condition description for its short-range term; and (ii) a Bonn--J\"{u}lich 
NNLO chiral EFT potential model \cite{KHM14} with a similar long-range 
behavior that is subject, however, to a strict power counting hierarchy, 
and in which the short-range behavior is given by suitably determined 
contact terms. The present work is not intended to compare between different 
microscopic ${\bar N}N$ potential models, nor to study possible ${\bar N}N$ 
quasi-bound states near threshold ({\it e.g.} with quantum numbers $^{11}S_0$ 
\cite{LWy05}, or $^{13}P_0$ \cite{EFe06}, or $^{31}S_0$ \cite{KHM15}), but 
rather to apply a microscopic model in the context of antiproton-nucleus 
interactions below threshold and at very low energies. As our handling 
of in-medium scattering amplitudes assumes some rather general properties 
of nuclei, we do not consider very light nuclei in the present work. 

${\bar N}N$ potentials are found to be strongly attractive and absorptive 
in all the microscopic models known to us, including the various versions 
of the Paris potential and the recently published potentials mentioned 
above. This results, generally, in {\it repulsive} ${\bar N}N$ $S$-wave 
scattering amplitudes at low energies. Hence, the simple impulse-approximation 
$t_{\bar p N}\rho$ optical potential, with $t_{\bar p N}$ the corresponding 
free-space $\bar p N$ $t$ matrix and $\rho$ the nuclear density, is repulsive 
at and near threshold. However, past global analyses of antiprotonic atoms 
\cite{FGM05,Fri14} achieved good agreement with experiment when using an 
empirical (as opposed to `microscopic') local {\it attractive} and absorptive 
optical potential, related to nuclear densities through a folded-in 
finite-range interaction of rms radius about 1.1--1.2~fm. A strong density 
dependence of the effective, in-medium $\bar p N$ $t$ matrix is apparently 
required in order to reverse the sign of the free-space $\bar p N$ $t$ matrix 
in the medium and inflate its size, or a significant contribution from 
$P$-wave amplitudes is able to achieve it. This problem was recognised 
already in the 1980s, with several many-body mechanisms suggested for 
obtaining an attractive low-energy $\bar p$-nucleus optical potential 
\cite{GWy82,SNa83,KGE84}, but none of these works was able to test such 
proposed mechanisms in actual global analyses of antiprotonic atoms
as the high-quality data of the PS209 experiment \cite{TJC01} 
were non-existent then. Previous 
attempts to add empirically a $P$-wave potential term or non linear density 
terms \cite{BFG95} were unsuccessful, failing among other things to respect 
constraints imposed by neutron density distributions \cite{FGa07}. More 
specifically, an imaginary part of a $P$-wave potential compatible with an 
earlier version of the Paris potential~\cite{CLL82} could be accommodated, 
but then the real part of the $S$-wave term was found to be incompatible with 
the Paris potential. It is therefore interesting to apply the present approach 
of treating energy and density dependence (reviewed in \cite{GFB14}) to the 
latest version of the Paris potential \cite{ELL09}. 

The paper is organized as follows. Section~\ref{sec:theory} is a brief 
description of the present approach. In Subsection~\ref{subsec:kin} we 
introduce the in-medium kinematics satisfied in hadron-nucleus collisions, 
in which the hadron-nucleon center of mass (cm) energy depends also on the 
{\it momenta} of the participating particles, and in 
Subsection~\ref{subsec:wrw} we discuss in-medium corrections to the free 
$\bar p N$ amplitudes such as Pauli correlations. Section \ref{sec:ampl} 
describes the free-space $S$-wave and $P$-wave ${\bar p}N$ input amplitudes 
derived from the 2009 version of the ${\bar N}N$ Paris potential and used 
in the present work. Low-energy ${\bar N}p$ annihilation is also considered 
in this section. Section \ref{sec:atoms} reports on results of comprehensive 
fits to strong-interaction observables in antiprotonic atoms, with special 
emphasis placed on the role played by the $P$-wave ${\bar p}N$ amplitudes 
in reproducing the main features of best-fit empirical $\bar p$-nucleus 
optical potentials. In Section \ref{sec:above} we present some results 
for the elastic scattering of 48 MeV antiprotons by C, Ca and Pb nuclei. 
Annihilation cross sections of antiprotons and antineutrons on nuclei are 
also briefly mentioned. Section \ref{sec:summ} offers a brief discussion 
and summary of the present study.

\section{Theoretical background}
\label{sec:theory}

Here we present the essentials of the theoretical background for the
present work, referring to previous publications for further details.

Strong interaction observables in pionic and kaonic atoms are usually
calculated \cite{FGa07} from the relativistic Klein-Gordon equation
\begin{equation}\label{eq:KG}
\left[ \nabla^2  - 2{\mu}(B+V_{\rm opt} + V_c) + (V_c+B)^2\right] \psi = 0,
\end{equation}
where $\hbar = c = 1$ is implicitly assumed.
Here $\mu$ is the meson-nucleus reduced mass, $B$ is the complex binding
energy, $V_c$ is the finite-size Coulomb interaction of the meson with
the nucleus, including vacuum-polarization terms. 
$V_{\rm opt}$ is the optical potential describing the strong interaction
of the meson with the nucleus. For antiprotonic
atoms we can use this equation because 
for a given $l$ in 
good approximation it gives the spin-averaged results of the Dirac
equation \cite{FGM05}. This is certainly acceptable with the 
experimental results
of the PS209 collaboration which forms the basis of the present work.

One of the aims of the present work is to elucidate the role of
the $P$-wave part of the antiproton-nucleon interaction in
antiproton-nucleus interactions near threshold \cite{BFG95,CLL82,PLL94}. 
 Therefore we include it explicitly in the optical potential, which we
take in analogy to the pion-nucleus potential \cite{FGa07}
\begin{equation} \label{eq:EE1}
2\mu_{\bar p} V_{\rm opt}(r) = q(r) + 3\vec \nabla \cdot \alpha(r) \vec \nabla,
\end{equation}
where $\mu _{\bar p}$ is the $\bar p$-nucleus reduced mass but 
unlike for pions, a factor $2l+1=3$ is 
introduced explicitly into the
$P$-wave part to match the normalization of the amplitudes of the
following Section. The $S$-wave part is written as
\begin{equation} \label{eq:EEs}
q(r)  =  -4\pi(1+\frac{\mu_{\bar p}}{m_N})\{{b}_0[\rho_n(r)+\rho_p(r)]
  +{b}_1[\rho_n(r)-\rho_p(r)] \}   
\end{equation}
and the $P$-wave part 
\begin{equation} \label{eq:EEp}
\alpha (r)  =  4\pi(1+\frac{\mu_{\bar p}}{m_N})^{-1} \{ c_0[\rho_n(r)+\rho_p(r)]
  +c_1[\rho_n(r)-\rho_p(r)] \},
\end{equation}
where $\rho_n(r)$ and $\rho_p(r)$ are the neutron and proton densities
 normalized to N and Z, respectively, with N+Z=A.
If evaluated at threshold the parameters $b_{0,1}$ and $c_{0,1}$ are
related to the scattering lengths and to the scattering volumes, respectively.
However, in the present work these parameters are evaluated at 
density-dependent energies as explained below.

\subsection{In-medium kinematics} 
\label{subsec:kin} 

The model underlying the subthreshold energy algorithm adopts the Mandelstam
variable $s = (E_{\bar p} + E_N)^2 -(\vec p _{\bar p} + \vec p_N)^2$ as the
argument transforming free-space to in-medium 
antiproton-nucleon amplitudes,
where both the $\bar p$ and the nucleon variables are determined independently
by the respective environment of a $\bar p$ atom and a nucleus. Consequently,
unlike in the two-body cm system, here $\vec p _{\bar p} + \vec p_N$ does not
vanish, and one gets to a good approximation
$(\vec p _{\bar p}+\vec p_N)^2\rightarrow p _{\bar p}^2+p_N^2$
upon averaging over angles. The energies are given by
\begin{equation}
\label{eq:energies}
E_{\bar p} =m_{\bar p} -B^R_{\bar p},~~~~ E_N=m_N-B_N,
\end{equation}
where $B^R_{\bar p}$ is the real part of the $\bar p$ binding energy 
in the atom, $B_N$ the (real) binding energy of the nucleon 
and $m$ are masses. 
For the $\bar p$ 
momentum we substitute locally
\begin{equation}
\label{eq:locpi}
\frac{p_{\bar p}^2}{2 m_{\bar p}} = -B^R_{\bar p} - {\rm Re}~V_{\rm opt} -V_c.
\end{equation}
For the nucleon we adopt the Fermi gas model (FGM), yielding in the local
density approximation
\begin{equation}
\label{eq:Fermi}
\frac{p_N^2}{2 m_N} = T_N\; (\rho / {\bar\rho})^{2/3},
\end{equation}
where $\rho$ is the local density, $\bar\rho$ is the average nuclear density 
and $T_N$ is the average nucleon kinetic energy which assumes the value 
23~MeV in the FGM.

Defining $\delta \sqrt s =\sqrt s -E_{\rm th}$ with 
$E_{\rm th}=m_{\bar p}+m_N$,
then to first order in $B/E_{\rm th}$ and $(p/E_{\rm th})^2$ one gets
\begin{equation}
\delta \sqrt s = -B_N\rho/{\bar \rho}
-\xi_N [T_N(\rho/\bar{\rho})^{2/3}+B^R_{\bar p}\rho/\rho_0]
 +\xi_{\bar p}[{\rm Re}~V_{\rm opt}+V_c (\rho/\rho_0)^{1/3}],
\label{eq:SCmodif}
\end{equation}
with $\xi_N=m_N/(m_N+m_{\bar p}),~\xi_{\bar p} = m_{\bar p}/(m_N+m_{\bar p})$,
and $\rho_0=0.17$~fm$^{-3}$. Following previous applications
\cite{CFG11,CFG11a,FGa12,GMa12,FGa13} an average binding energy value of
$B_N=8.5$~MeV is used. The specific $\rho/\rho_0$ and $\rho/{\bar \rho}$
forms of density dependence ensure that $\delta\sqrt{s}\rightarrow 0$ when
$\rho \rightarrow 0$~\cite{FGa13}.

Another variant of Eq.~(\ref{eq:SCmodif}) is obtained when considering the
minimal substitution requirement, the importance of which for incorporating
electromagnetism in a gauge-invariant way into the pion optical potential
was first pointed out by Ericson and Tauscher~\cite{ET82} and more recently
emphasized by Kolomeitsev, Kaiser and Weise~\cite{KKW03}. Indeed, the
application of minimal substitution has been successful in analyses of pionic
atoms \cite{FGa07,FGa04} and pion scattering at 
low energies \cite{FGa14,Fri05}.
Here $E=E_{\bar p}+E_N$ is replaced by $E-V_c$ and 
then Eq.~(\ref{eq:SCmodif}) becomes
\begin{equation}
\delta \sqrt s =
-B_N\rho/{\bar \rho}
-\xi_N [T_N(\rho/\bar{\rho})^{2/3}+B^R_{\bar p}\rho/\rho_0+
V_c (\rho/\rho_0)^{1/3}]
+\xi_{\bar p}{\rm Re}~V_{\rm opt}.
\label{eq:SCmodifMS}
\end{equation}
Eq.~(\ref{eq:SCmodifMS}) is used in the present work to handle the in-medium 
kinematics in antiprotonic atoms. At energies above threshold, in applications 
to scattering and in-flight annihilation, the term $-\xi_{N} B^{R}_{\bar p}
\rho/\rho_0$ is replaced by $\xi_{N}E_{\rm lab}$, with $E_{\rm lab}$ the beam 
kinetic energy, leading to 
\begin{equation}
\delta \sqrt s =
-B_N\rho/{\bar \rho}
-\xi_N [T_N(\rho/\bar{\rho})^{2/3}-E_{\rm lab}+V_c (\rho/\rho_0)^{1/3}]
+\xi_{\bar p}{\rm Re}~V_{\rm opt}.
\label{eq:SCmodifMSrev}
\end{equation}

\subsection{In-medium amplitudes} 
\label{subsec:wrw} 

To obtain in-medium amplitudes from the free space ones we
apply the multiple scattering approach of Waas, Rho and Weise (WRW)
\cite{WRW97} as used for kaonic atoms in Ref.~\cite{FGa13}. Since
the isospin structure of $\bar K N$ and $\bar p N$ is the same,
we use for the $S$-wave potential $q(r)$ of Eq.~(\ref{eq:EEs}) 
the form given in Ref.~\cite{FGa13},
\begin{equation}
q(r)=-4\pi\left[ \frac{(2\tilde f_{\bar pp}- 
\tilde f_{\bar pn})\:\frac{1}{2}\rho_p}{1+\frac{1}{4}\xi_{k=0}(
2\tilde f_{\bar pp}-
 \tilde f_{\bar pn})
\rho(r)}+\frac{\tilde f_{\bar pn}(\frac{1}{2}\rho_p+\rho_n)}
{1+\frac{1}{4}\xi_{k=0} \tilde f_{\bar pn}\rho(r)}\right]\;,
\label{eq:WRW}
\end{equation}
where $\rho=\rho_p +\rho_n$ and $\tilde f=\zeta f$ is a $\bar p$-nucleus cm 
amplitude related to the two-body cm amplitude $f$, with $\zeta $ given by
\begin{equation}
\zeta=1+\frac{A-1}{A}\frac{\mu_{\bar p}}{m_N}.
\label{eq:kinfac}
\end{equation}
To leading order, $\xi_{k=0}=9\pi/p_F^2$ with $p_F$ the local Fermi momentum, 
accounting for nuclear Pauli correlations. These WRW medium corrections apply 
only to the $S$-wave part of the potential~\cite{WRW97}. The $P$-wave part, 
Eq.~(\ref{eq:EEp}), is taken from the free amplitudes at energies prescribed 
by Eq.~(\ref{eq:SCmodifMS}) for antiprotonic atom applications, or by 
Eq.~(\ref{eq:SCmodifMSrev}) for applications above threshold, without 
incorporating further nuclear correlations such as realized in pionic atoms 
by the Lorentz-Lorenz modifications. This is justified in $\bar p$ atoms 
owing to the extremely low nuclear densities encountered in their analysis.

\begin{figure}[htb]
\begin{center}
\includegraphics[height=75mm,width=0.75\textwidth]{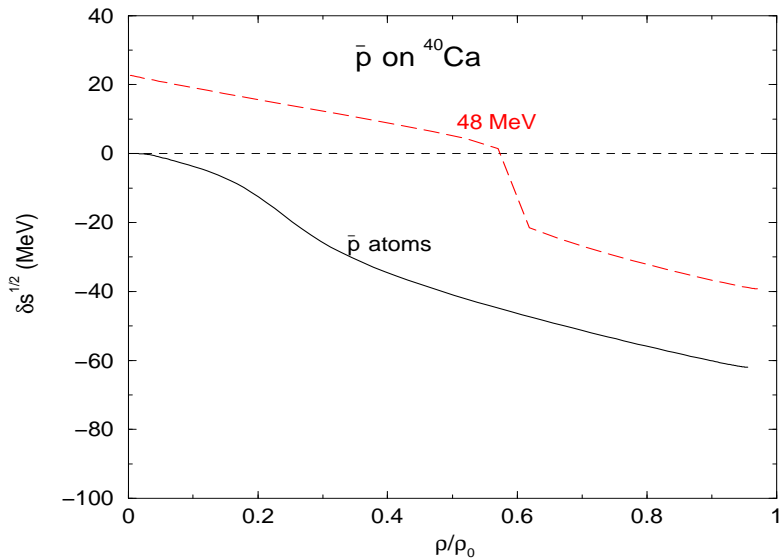}
\caption{Examples of the density to energy transformation from
Eqs.~(\ref{eq:SCmodifMS}) and (\ref{eq:SCmodifMSrev}), 
using Eq.~(\ref{eq:WRW}).} 
\label{fig:deltavsrhoR} 
\end{center}
\end{figure}

Equations (\ref{eq:SCmodifMS}) and (\ref{eq:SCmodifMSrev}) define a density 
to energy transformation through the scattering amplitudes used in the 
calculation of antiproton-nucleus optical potential Eq.~(\ref{eq:WRW}). 
As the real part of the potential determines the energy and, in turn, 
the energy and density determine the amplitudes, a self-consistent solution 
is required. 
Examples for this transformation are shown in Fig.~\ref{fig:deltavsrhoR} 
for $\bar p$ atoms of $^{40}$Ca and for elastic scattering of 48 MeV 
$\bar p$ using the amplitudes from the 2009 version of the Paris $\bar NN$ 
potential. The energy shift $\delta \sqrt s$ for $\bar p$ atoms comes out 
negative definite over the full density range considered, reaching fairly 
large values of up to about $-60$~MeV at nuclear-matter density $\rho_0$, 
in agreement with its behavior for kaonic atoms~\cite{FGa13}. For low-energy 
scattering, in contrast, $\delta \sqrt s$ changes sign as a function of the 
density from positive values to negative ones around $0.6 \rho_0$, and its 
slope exhibits a marked discontinuity caused by the in-medium $S$-wave 
amplitude, Eq.~(\ref{eq:WRW}), switching from repulsion to attraction as 
the  density is increased. In the case of $\bar p$ atoms, where the input 
$\bar p N$ $S$-wave amplitudes are appreciably larger in size than for 
48~MeV antiprotons, the transition from repulsion to attraction takes 
place at very small densities, thus making it unobservable on the 
$\delta s^{1/2}$ curve. Further discussion of the WRW mechanism 
responsible for the in-medium sign reversal of the real part of the 
$S$-wave potential $q(r)$, Eq.~(\ref{eq:WRW}), is deferred to Appendix A, 
including some realistic estimates.

\section{Free-space input amplitudes}
\label{sec:ampl}
\subsection{General}
\label{subsec:genaral}

\begin{figure}[htb]
\begin{center}
\includegraphics[height=75mm,width=0.75\textwidth]{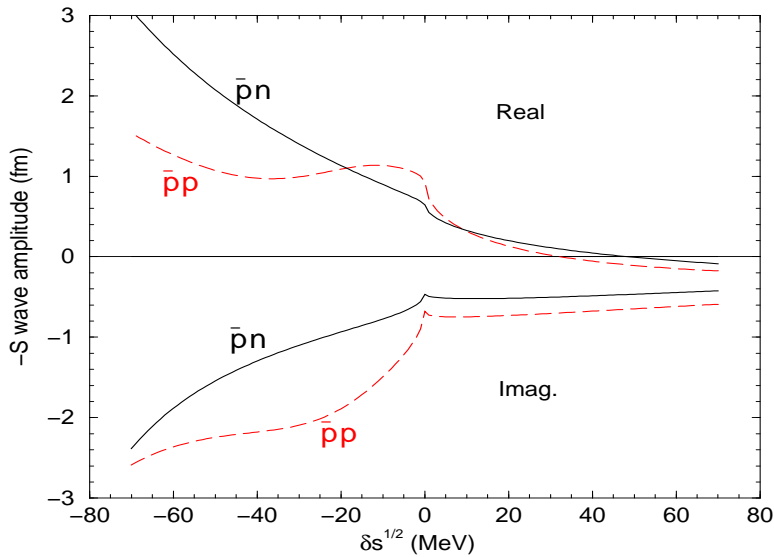} 
\caption{Antiproton-nucleon $S$-wave scattering amplitudes around 
threshold from the 2009 Paris potential \cite{ELL09}. Negative values 
represent attraction and absorption.} 
\label{fig:swave} 
\end{center} 
\end{figure} 

\begin{figure}[htb]
\begin{center}
\includegraphics[height=75mm,width=0.75\textwidth]{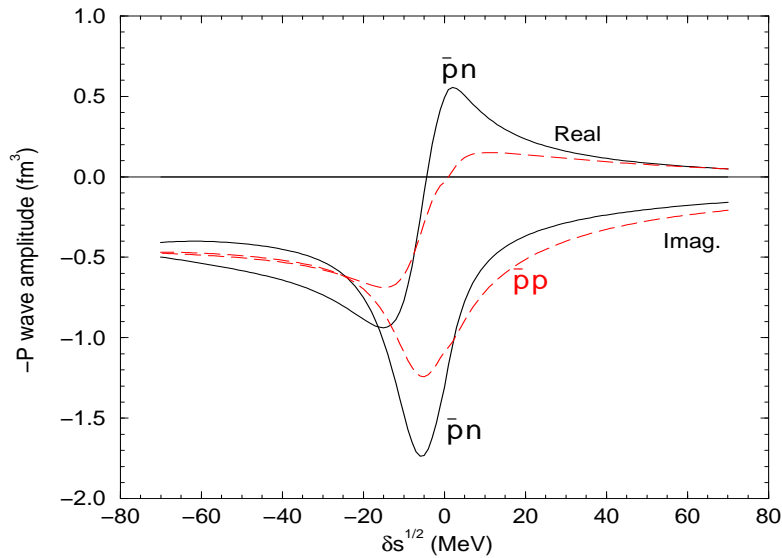}
\caption{Antiproton-nucleon $P$-wave scattering amplitudes around 
threshold from the 2009 Paris potential \cite{ELL09}.
Negative values represent attraction and absorption}
\label{fig:pwave}
\end{center}
\end{figure}

Here we briefly describe the free-space ${\bar p}N$ amplitudes derived
from the 2009 version of the ${\bar N}N$ Paris potential \cite{ELL09}. 
It was suggested by Green and Wycech \cite{GWy82} that the ${\bar p}N$ 
amplitudes most appropriate for use in antiprotonic atoms are half 
off-shell, $\langle {\vec p}~|f(E)|{\vec {p'}}\rangle$ with one on-shell 
cm momentum $p~=~\sqrt{m_NE}$ and one off-shell cm momentum which we choose 
as $p'=0$ to focus on near-threshold energies. Figs.~\ref{fig:swave} and 
\ref{fig:pwave} show such $S$-wave and $P$-wave half off-shell ${\bar p}N$ 
amplitudes, respectively, derived from the 2009 version of the ${\bar N}N$ 
Paris potential. The ${\bar p}n$ amplitudes are pure isospin $T=1$ ${\bar N}N$ 
amplitudes, whereas the ${\bar p}p$ amplitudes are equal-weight mixtures 
of isospins $T=0$ and $T=1$ ${\bar N}N$ amplitudes. Apart from their isospin 
structure, these amplitudes represent angular-momentum averages, appropriate 
for use in antiprotonic atoms, over states denoted by $^{(2T+1)(2S+1)}L_J$ 
where $J,S,L$ are the total, spin and orbital angular momentum, respectively. 
The actual $\bar p p$ and $\bar p n$ $S$-wave ($L=0$) and $P$-wave ($L=1$) 
amplitudes plotted in Figs.~\ref{fig:swave} and \ref{fig:pwave}, respectively, 
are derived from the $T=0$ and $T=1$ ${\bar N}N$ pure isospin amplitudes, 
which are obtained by the appropriate angular-momentum averaging of fixed-$T$ 
amplitudes as follows
\begin{equation} 
\label{eq:statistical} 
f^{T}_{L} = \frac{~\sum_{SJ}~(2J+1)~f^{TJ}_{LS}~}{\sum_{SJ}~(2J+1)}. 
\end{equation}
 
The real part of the Paris potential is predominantly attractive and 
generates quasi-bound states or resonances in specific partial waves. 
However, in the $S$-wave amplitudes these are washed out by the very 
strong absorption which acts repulsively. In particular the 2009 
version of this potential  generates a broad quasi-bound 
$^{11}S_0$ state at $E = -4.8 -i26$ MeV. This state has a small 
statistical weight and it causes a small anomaly just below threshold, 
producing no effect on ${\bar p}$-nuclear observables near threshold.  
It is seen from Fig.~\ref{fig:swave} that the free $S$-wave 
amplitudes, in distinction from the underlying attractive potential, 
represent repulsion over a very wide energy range. 
This disagrees with the empirical attractive potential 
deduced from previous analyses of antiprotonic atoms unless 
medium effects, such as the WRW prescription discussed 
in Section \ref{sec:theory} or other amplitudes, e.g. $P$-wave, 
produce attraction.

Fig.~\ref{fig:pwave} shows the $P$-wave amplitudes from the Paris
potential. In the $P$-wave states the short range annihilation 
has much weaker effect and the attractive nature of the potential 
is evident. In particular a quasi-bound, fairly narrow state 
at $E = -4.5 -i9.0$ MeV is generated in the $ ^{33}P_1$ partial wave. 
Due to its sizable statistical weight it is seen clearly in the figure. 
In contrast to the $S$-wave quasi-bound state discussed above the $P$ 
level is robust, arising in all versions of the Paris potential. In the 
2009 version its position is rather reliably fixed by the $\bar p p$ 
scattering volumes extracted from the antiprotonic hydrogen atom.

\begin{figure}
\begin{center}
\includegraphics[height=75mm,width=0.75\textwidth]{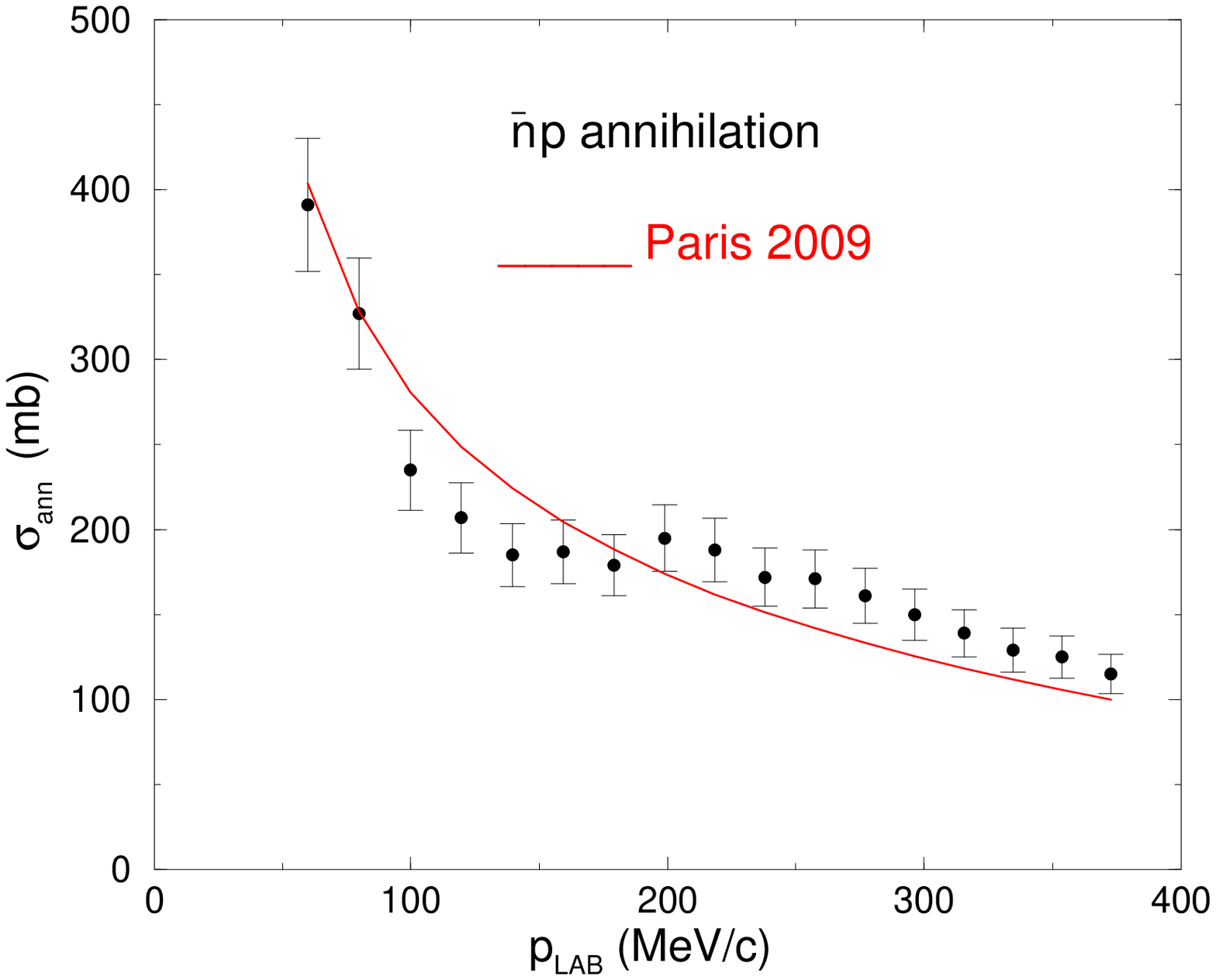}
\caption{Comparison between predictions and
experiment \cite{Ber97} for $\bar np$ annihilation.}
\label{fig:nbarann}
\end{center}
\end{figure}

\subsection{Low energy antinucleon-proton annihilation}
\label{subsec:nbarN}

\begin{figure}
\begin{center}
\includegraphics[height=75mm,width=0.75\textwidth]{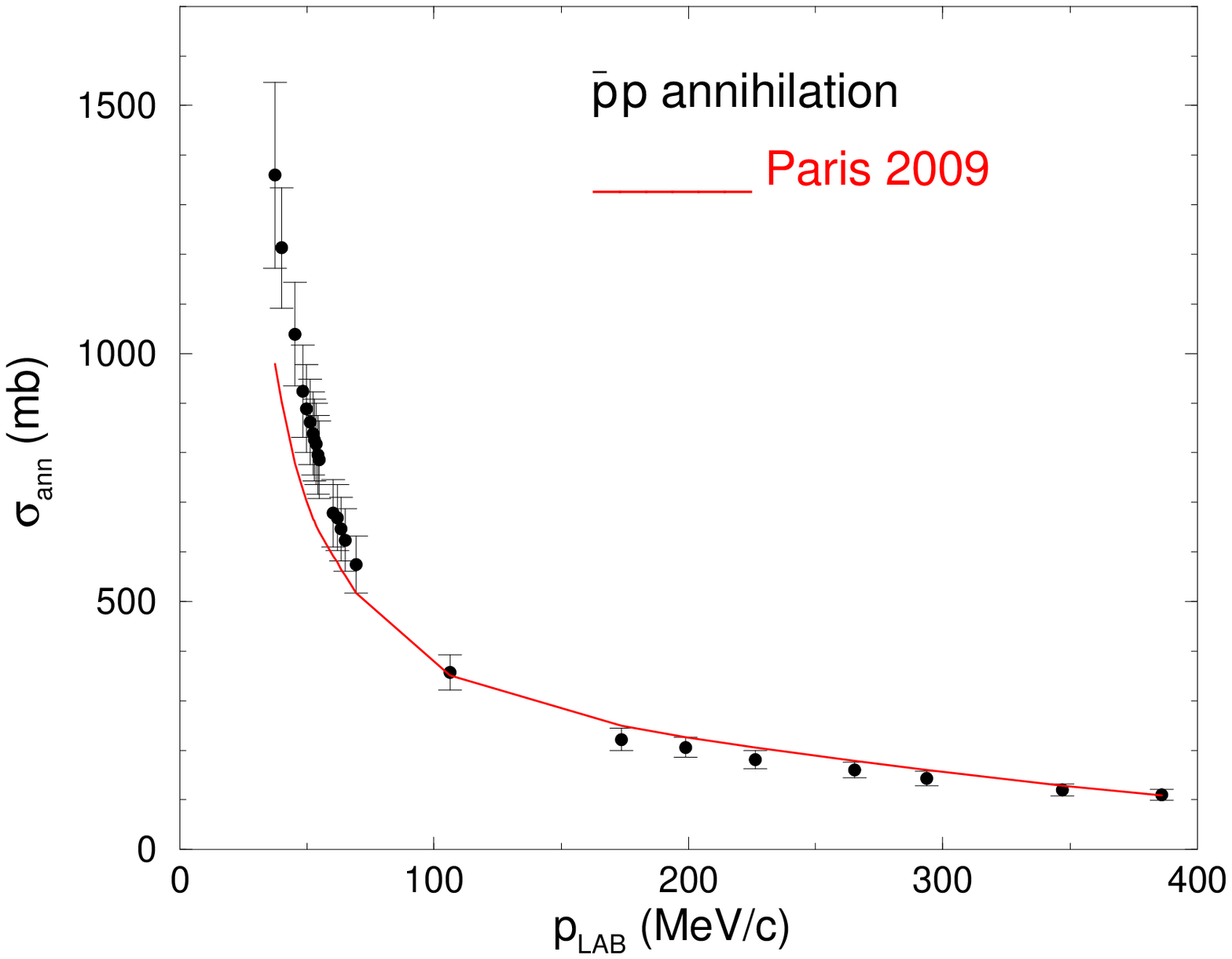}
\caption{Comparison between predictions and
experiment \cite{BBC96,ZBB99,Ben97,BCD90} for $\bar pp$ annihilation.}
\label{fig:pbarann}
\end{center}
\end{figure}

Before applying the Paris 2009 amplitudes to the $\bar p$-nucleus interaction
close to threshold we examine how well these amplitudes describe experimental
results for the $\bar N$-proton interaction at very low energies.

The obvious starting point is the strong interaction 
shift and width observed in $\bar p$H
atoms \cite{GAA99,Got04}  
which, with the help of the Deser  \cite{DGB54} or Trueman equations
\cite{Tru61} approximates the
scattering length for the $\bar pp$ interaction. Using a modern version
of the latter \cite{MRR06} we have calculated also
the scattering length in the
absence of the Coulomb interaction using Eq.~(20) of Ref.~\cite{BFG01}.
The result of $-$0.81+$i$0.72 fm is in reasonable agreement with the 
amplitudes used in the present work.

Next we compare predictions of in-flight annihilation cross sections up 
to 400~MeV/c with calculations based on the $S$- and $P$-wave amplitudes. 
Figure~\ref{fig:nbarann} shows $\bar n p$ annihilation cross sections from 
the OBELIX collaboration \cite{Ber97}. As some of the quoted experimental 
errors are unrealistically small, we have assumed errors of $\pm$10\% at 
all points. No parameter adjustment has been made in the calculation. 
Except for the small bump near 200 MeV/c the agreement is very satisfactory. 
Fig.~\ref{fig:pbarann} shows similar results for $\bar pp$ annihilation 
\cite{BBC96,ZBB99,Ben97,BCD90}. A subtle point may arise here because the 
2009 amplitudes were calculated without the Coulomb interaction. 
An approximate correction has therefore been applied by multiplying the 
$S$-wave and $P$-wave amplitudes by the Coulomb phase correction, 
namely,~$e^{2i\sigma_0}$~and $e^{2i\sigma_1}$, respectively, where $\sigma$ 
are the Coulomb phases. At the lowest momentum  the calculated cross section 
is then increased by  28\%, whereas near 400 MeV/c this correction is less 
than 2\%. The overall agreement between predictions and experiment is 
certainly acceptable.

\section{Antiprotonic atoms}
\label{sec:atoms}

The PS209 data used in the present work consist of strong interaction 
observables for 27 nuclear targets from $^{40}$Ca to $^{208}$Pb, 
totaling 84 data points \cite{TJC01}. Six points from earlier 
measurements on $^{16,18}$O \cite{KBB86} were also included.
Typical examples of results for various choices of potentials are shown 
in Table~\ref{tab:res}. Values of $\chi ^2$ for the 90 points indicate
the quality of fit, obtained when varying 2-3 parameters. Where
appropriate, the present results are in full agreement with earlier
analyses \cite{FGM05,Fri14}.

The first row of Table \ref{tab:res} shows, as a reference, the best fit 
obtained with a local empirical attractive and absorptive potential, 
including finite-range folding \cite{FGM05,Fri14}. The very strong 
absorption of antiprotons in nuclei confines the interaction with the 
atomic $\bar p$ to the surface of the nucleus \cite{FGa07} and this is 
demonstrated in the second row of the table, where we used only an empirical, 
energy independent $P$-wave potential. It is evident that equally good fits 
can be obtained when the potential is centered near the surface, where the 
gradient terms are effective. The third row is for a potential constructed 
from the free-space Paris 2009 amplitudes of the previous section, in which 
the WRW in-medium modification (\ref{eq:WRW}) is incorporated with in-medium 
kinematics satisfying the $\delta \sqrt s$ algorithm for $\bar p$ atoms, 
Eq.~(\ref{eq:SCmodifMS}). No adjustable parameters are included. It is clear 
from the value of $\chi ^2$ that no agreement with experiment is possible. 
The reasons for this failure are further discussed below in the last 
paragraph of the present section. 

\begin{table}[hbt]
\caption{Comparisons between calculation and experiment for $\bar p$ atoms, 
using various options of $S$-wave and $P$-wave potentials. The symbol '09 
means that $\bar pN$ amplitudes were used over a range of $\sqrt s$ values 
as given by Eq.~(\ref{eq:SCmodifMS}), including the WRW modification 
Eq.~(\ref{eq:WRW}). When indicated, these were multiplied by a scaling 
factor given in parentheses. When units of fm$^3$ are listed, the 
corresponding parameter was empirical. See text for more details.} 
\label{tab:res} 
\begin{center} 
\begin{tabular}{ccccc} 
\hline 
  & $S$-wave&Real $P$-wave&Imag. $P$-wave&$\chi^2(90)$ \\ \hline
1 &emp.&-&-&199  \\
2 & -&1.9$\pm$0.1 fm$^3$&2.8$\pm$0.1 fm$^3$& 206 \\
3 &  '09 & '09 &'09 & 2304  \\
4 &'09&2.9$\pm$0.1 fm$^3$&1.8$\pm$0.1 fm$^3$&203\\
5 &'09&'09$\times$($-$10.0$\pm$0.9)&'09$\times$(3.1$\pm$0.2)&571\\
6 &'09&2.9$\pm$0.1 fm$^3$&'09$\times$(1.3$\pm$0.1)&218\\
\hline
\end{tabular}
\end{center}
\end{table}

The fourth row of Table~\ref{tab:res} shows the results obtained by 
replacing the free-space $P$-wave term by an empirical one, while 
retaining the $S$-wave part as above, namely, the medium-modified 
amplitudes at density-dependent energies given in the present model. 
The fit to the data, varying two parameters, is as good as the best one. 
Comparing parameters with the results on the second row, we see that 
the real $P$-wave part is more attractive now as it has to overcome the 
repulsion of the $S$-wave part near the surface. The opposite is true 
for the imaginary part because the imaginary part of the $S$-wave term  
already provides some of the absorption. 

Row five of the table shows an attempt to modify the free-space $P$-wave 
amplitudes by a factor, separately for the real and imaginary potentials, 
within the present full approach. This was not successful as is seen from 
the value of $\chi ^2$ and by the scaling factor imposed on the real part 
which requires a sign change and an order of magnitude enhancement. 
Realizing that the major difficulty with the free-space amplitudes is 
probably with the real part of the $P$-wave, the last row shows a very 
good fit to the data when rescaling the imaginary part of the $P$-wave 
amplitude by 30\% while using an empirical real part. The latter agrees 
well with the one in row 4. 

\begin{figure}
\begin{center}
\includegraphics[height=70mm,width=0.75\textwidth]{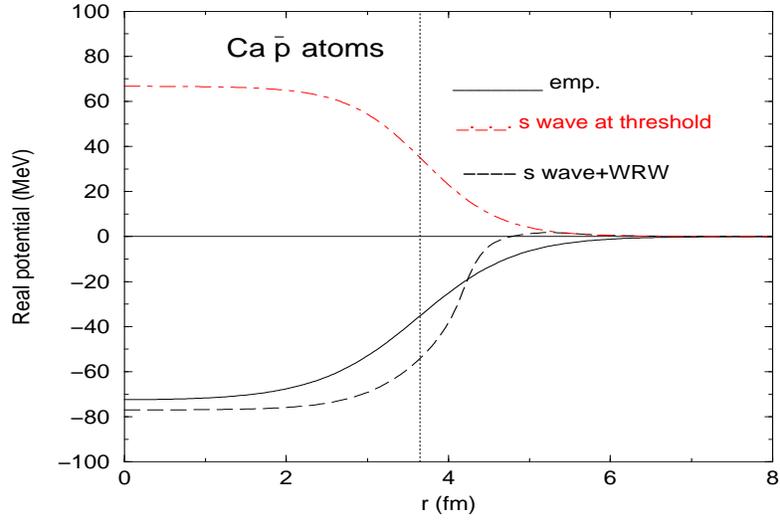}
\caption{Real part of antiproton-nucleus potentials for $^{40}$Ca $\bar p$ 
atoms. Solid curve for the best-fit empirical potential, dot-dash curve 
for the unmodified $S$-wave potential at threshold, dashed curve for the 
medium-modified $S$-wave potential. Vertical dotted line marks the 
half-density radius of $^{40}$Ca.}
\label{fig:Caatreal}
\end{center}
\end{figure}

\begin{figure}
\begin{center} 
\includegraphics[height=70mm,width=0.75\textwidth]{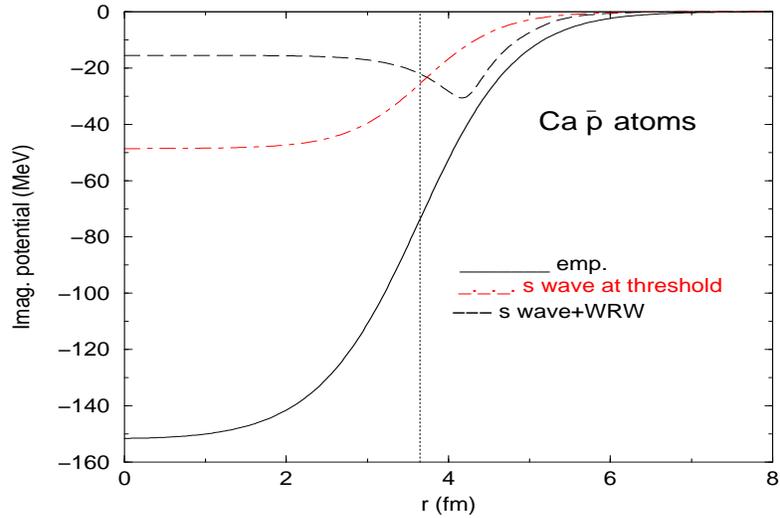} 
\caption{Imaginary part of antiproton-nucleus potentials for $^{40}$Ca 
$\bar p$ atoms, see caption to Fig.~\ref{fig:Caatreal} for the meaning of 
the various curves.} 
\label{fig:Caatimag} 
\end{center} 
\end{figure} 

Figures~\ref{fig:Caatreal} and \ref{fig:Caatimag} show real and imaginary 
parts of the $\bar p$-$^{40}$Ca potentials, respectively, typical of the 
results for the whole data base. Solid curves show the best-fit empirical 
potential which is attractive and obviously absorptive. The repulsive real 
part obtained from the free-space amplitudes at threshold is also shown 
(dot-dash) and the effects of the medium modification are clearly seen on the 
dashed curves. In the medium the WRW modification, assisted by the negative 
$\delta \sqrt{s}$ shift for $\bar p$ atoms (see Fig.~\ref{fig:deltavsrhoR}), 
turns the repulsion into attraction as demonstrated in Appendix A, but far 
down the surface the effect disappears and the potential becomes again 
repulsive, differing significantly from the empirical potential that 
fits well the data. Regarding the in-medium imaginary potential, it falls 
short of its empirical counterpart in the surface region. The WRW medium 
modification reduces the imaginary potential in the interior by a factor 
about three while enhancing it in the surface, but insufficiently to match 
the best-fit potential. The vertical dotted line in both figures indicates 
the half-density radius of the nucleus, thus demonstrating that the relevant 
region is at larger radii. With these observations it is clear why the 
application of the medium-modified Paris potential amplitudes to antiprotonic 
atoms fails, as was shown in the third row of Table~\ref{tab:res}.

\section{$\bar N$-nucleus above threshold} 
\label{sec:above} 

Calculated in-flight annihilation cross sections for $\bar p$ on light muclei, 
using the present in-medium procedures, are in good agreement with the 
very few available experimental results \cite{BBB86,BBB00}. However, 
the best test of a potential is by comparing predictions to measured 
differential cross sections. Although at 48 MeV beam energy $S$ and $P$ waves 
might be insufficient to fully describe {\it microscopically} the interaction 
of $\bar p$ with medium weight and heavy nuclei, we report below some features 
observed using the Paris 2009 potential at that energy.

\subsection{Scattering of 48 MeV antiprotons}
\label{subsec:scatt}

Measurements of elastic scattering of antiprotons by $^{12}$C, $^{40}$Ca and 
$^{208}$Pb were made in the 1980s and analyzed using standard low-energy 
optical model methods, see Janouin et al. \cite{JLG86} and references therein. 
Here we apply the approach used above for antiprotonic atoms also to the 
$\bar p$-nucleus interaction at 48~MeV energy. In parallel with $\bar p$ 
atoms, excellent fit to the data could be obtained with an empirical local 
potential, including finite-range folding with rms radius of 1.5$\pm$0.1 fm, 
as is summarized in the first row of Table~\ref{tab:res48}. 
This range is somewhat larger than the corresponding one for $\bar p$ atoms. 
An attempt to use only an empirical $P$-wave potential was not successful 
unless finite-range folding was introduced, unlike with atoms, and then 
only a moderate fit was obtained, see the second row of the table. 
Very good fits were possible by a combination of empirical local and 
$P$-wave potentials, but then correlations prevented achieving a unique 
solution. Experience shows that no more than three meaningful parameters 
can be derived here. 

\begin{table}[hbt]
\caption{Comparisons between calculation and experiment for 48 MeV $\bar p$ 
scattering by $^{12}$C, $^{40}$Ca and $^{208}$Pb, using various options 
of $S$-wave and $P$-wave potentials. The symbol '09 means that $\bar pN$ 
amplitudes were used over a range of $\sqrt s$ values as given by 
Eq.~(\ref{eq:SCmodifMSrev}), including the WRW modification 
Eq.~(\ref{eq:WRW}). When indicated, these were multiplied by a scaling factor 
given in parentheses. When units of fm$^3$ are listed, the corresponding 
parameter was empirical. The last column lists the rms radii of the 
finite-range folding applied to the adjusted terms. See text for more 
details.} 
\label{tab:res48} 
\begin{center}
\begin{tabular}{cccccc}
\hline
  & $S$-wave&Real $P$-wave&Imag. $P$-wave&$\chi^2(83)$& range(fm)\\ \hline
1 &emp.&-&-&151 &1.5$\pm$0.1 \\
2 & -&0.5$\pm$0.05 fm$^3$&0.8$\pm$0.05 fm$^3$& 201 & 0.9$\pm$0.1\\
3 &  '09 & '09 &'09 & 1823 & -  \\
4 &'09&0.31$\pm$0.03 fm$^3$&0.50$\pm$0.03 fm$^3$&176 & 1.5$\pm$0.1 \\
5 &'09&'09$\times$($-$2.1$\pm$0.2)&'09$\times$(2.4$\pm$0.2)&495  & - \\
6 & '09 &'09$\times(-1.92\pm0.04)$&'09$\times$(1.32$\pm$0.09)&218 & 
1.5$\pm$0.1 \\
\hline
\end{tabular}
\end{center}
\end{table}

The third row of Table \ref{tab:res48} is for using in-medium amplitudes 
generated from free-space amplitudes of the Paris potential by using 
in-medium kinematics, Eq.~(\ref{eq:SCmodifMSrev}), and applying the WRW 
modification (\ref{eq:WRW}) to the $S$-wave part without any adjustable 
parameters. Clearly that is unacceptable, but row 4 shows a good fit when 
the 2009 energy dependent Paris-potential $P$-wave amplitudes are replaced 
by an empirical $P$-wave term. Row 5 is for attempts to fit the data by 
applying scaling factors to the Paris potential $P$-wave amplitudes. 
No fit is possible but we note that the real part required a change of sign, 
as is the case with atoms (see Table~\ref{tab:res}). Finally, row 6 shows 
that an almost acceptable fit is possible with finite range applied to the 
$P$-wave term but with significant scaling factors.

\begin{figure}
\begin{center}
\includegraphics[height=70mm,width=0.75\textwidth]{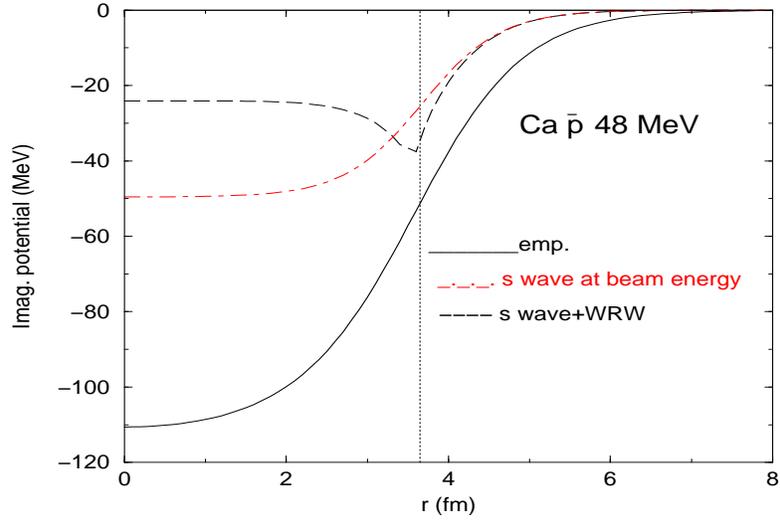}
\caption{Imaginary part of antiproton-nucleus potentials for $^{40}$Ca at
48 MeV. Solid curve for the best-fit empirical potential, dot-dash curve
for the unmodified $S$-wave potential at 48 MeV, dashed curve for the
medium-modified $S$-wave potential. Vertical dotted line marks the
half-density radius of $^{40}$Ca.}
\label{fig:Ca48imag}
\end{center}
\end{figure}

\begin{figure}
\begin{center} 
\includegraphics[height=70mm,width=0.75\textwidth]{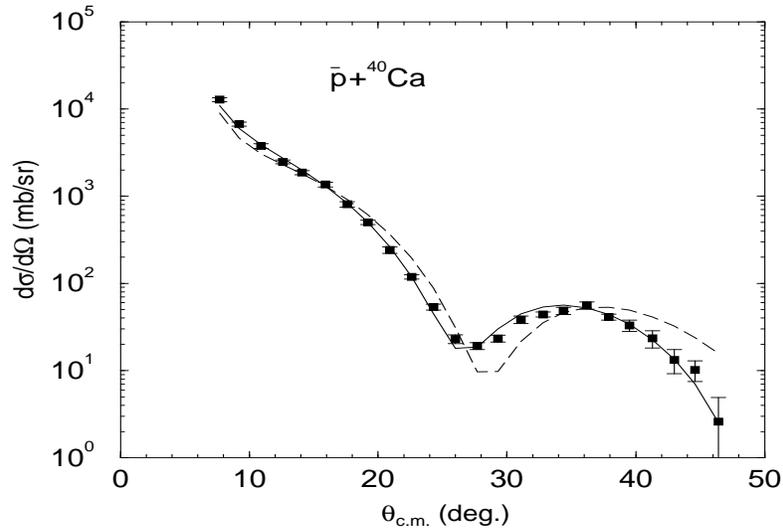} 
\caption{Differential cross sections for elastic scattering of 47.8 MeV 
$\bar p$ by $^{40}$Ca \cite{JLG86}. Solid curve for the best-fit empirical 
potential (row 1 of Table \ref{tab:res48}), dashed curve for 2009 Paris 
potential (row 3 of Table \ref{tab:res48}).}
\label{fig:Ca48diff}
\end{center}
\end{figure}

Figure~\ref{fig:Ca48imag} shows, as an example, the imaginary part of 
$\bar p$ nuclear potentials for 48 MeV antiprotons interacting with 
$^{40}$Ca. As for $\bar p$ atoms, there are significant differences 
in the surface region between the in-medium microscopic $S$-wave potential 
and the best-fit empirical one. Figure~\ref{fig:Ca48diff} compares 
experimental differential cross sections for elastic scattering of 
48 MeV $\bar p$ from $^{40}$Ca with calculations made with the best-fit 
empirical potential (row 1 of Table~\ref{tab:res48}) and with the in-medium 
microscopic potential obtained from the Paris 2009 free-space amplitudes 
(row 3 of the table). The poor agreement near the first minimum indicates 
that the microscopic potential is inadequate near the surface region of 
the nucleus. Indeed it is seen from Fig.~\ref{fig:pwave} that whereas it 
might be possible for the $P$-wave part to add the extra absorption required, 
there is no chance that the real part of the $P$-wave amplitude will close 
the gap between empirical potentials and the in-medium microscopic potentials 
generated from the free-space amplitudes of the Paris potential. 
That is also the conclusion reached from inspection of Table~\ref{tab:res48}. 
However, it must be emphasized that we have retained all along the $S$-wave 
amplitudes as given by the Paris potential. Therefore the difficulties with 
the $P$-wave part essentially mean that there are some inconsistencies 
between the two types of amplitudes, within the present model of handling 
amplitudes in the nuclear medium. A need to include a $D$ wave above 
threshold in a microscopic model cannot be ruled out.

\subsection{Annihilation on nuclei}
\label{subsec:annA}

There has been only a handful of measurements of annihilation cross sections 
of antiprotons on medium-weight and heavy nuclei at energies close to 
threshold, where the present optical model approach is being tested. 
In contrast, measurements of annihilation cross sections of antineutrons 
on nuclei across the periodic table at seven momenta between 76 and 375 
MeV/c were made by Astrua et al. \cite{ABB02}. These results have been 
compared \cite{Fri14,Fri15} with predictions by {\it empirical} optical 
potentials that fit quite well the few available $\bar p$-nucleus 
annihilation cross sections. Here we compare these results with predictions 
by potentials based on the Paris 2009 amplitudes.

\begin{figure}[htb]
\begin{center}
\includegraphics[height=75mm,width=0.75\textwidth]{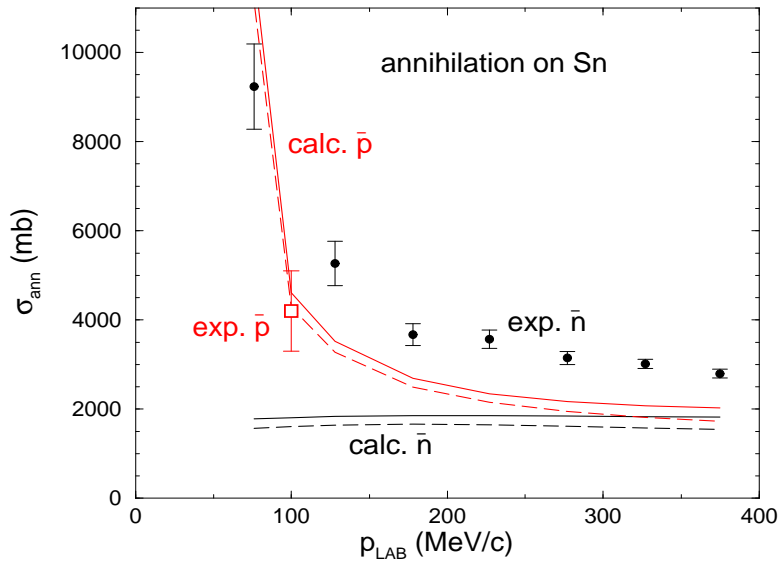}
\caption{Comparisons between calculated and experimental annihilation
cross sections on Sn. Solid circles for $\bar n$ \cite{ABB02}, open square 
for $\bar p$ \cite{BCH11}. Solid lines are for using in-medium kinematics, 
Eq.~(\ref{eq:SCmodifMSrev}), and applying the WRW modification 
Eq.~(\ref{eq:WRW}). Dashed lines are for $\delta\sqrt{s}=\xi_N E_{\rm lab}$.} 
\label{fig:SnJune15} 
\end{center} 
\end{figure} 

Figure~\ref{fig:SnJune15} shows, as an example, experimental 
annihilation cross sections for antineutrons on Sn (solid circles) 
and a single point (open square) for annihilation of antiprotons 
on the same target \cite{BCH11}. Calculations are shown as solid 
and dashed lines for the full $\delta\sqrt s$ model and for 
$\delta\sqrt s=\xi_N E_{\rm lab}$, respectively. 
The weak sensitivity to the model is due to the interaction being 
confined to the extreme surface region of the nucleus. The agreement 
with experiment for the single $\bar p$ point is very good. The sharp 
disagreement for the antineutrons was dicussed in \cite{Fri14,Fri15}, 
based on empirical optical potentials. It persists also here when 
the potential is constructed from the Paris 2009 amplitudes. Very recently 
it was suggested by Bianconi~et al.~\cite{BRM14}, using qualitative 
arguments, that annihilation cross sections could be enhanced by the 
interaction between the Coulomb field of a nucleus and an induced electric 
dipole of the antineutron. We have tested this idea using a realistic 
polarizability of 0.001 fm$^3$ \cite{BSc14} for the antineutron but the 
effect is totally negligible. Increasing the polarizability by an order 
of magnitude will lead to effects smaller than 1\%.

\section{Discussion and summary}
\label{sec:summ}

The ability of a simple empirical optical model approach to describe well
the interaction of sub-threshold and of low energy antiprotons with nuclei
has been known for some time. Among other things, this success could
result from the interaction being confined to the extreme surface
region of the nucleus due to the very strong absorption of $\bar p$
and $\bar n$ in nuclear matter. Nevertheless, it was interesting to
see how well can more microscopic approaches do in this respect.
This has been done in the present work using the 
2009 Paris $\bar NN$ potential \cite{ELL09}.

Inspection of optical potentials based on a recent algorithm for handling 
microscopic scattering amplitudes in the nuclear medium shows that it is 
necessary to include contributions from the $P$-wave part of the ${\bar p}N$ 
interaction, as suggested previously \cite{GWy82}. In the present case 
this is necessary because the $S$-wave amplitudes are repulsive throughout 
the full energy range, as seen in Fig.~\ref{fig:swave}, and although the 
WRW medium modification makes them attractive in the interior, they are 
nevertheless repulsive at the relevant low density region of the nucleus. 
This applies both to $\bar p$ atoms and to scattering at 48~MeV beam 
energy. The reversal of sign of the real potential 
in a sufficiently dense matter is exclusively due to the in-medium WRW 
modification (\ref{eq:WRW}) and not due to the in-medium kinematics algorithm, 
Eqs.~(\ref{eq:SCmodifMS}) and (\ref{eq:SCmodifMSrev}). Tests show that this 
phenomenon is linked to the particularly large values of both real and 
imaginary part of the $S$-wave amplitudes and it disappears when the 
amplitudes are reduced by a factor 3 or more. In practice we find that 
the real part of the $P$-wave amplitudes cannot substitute for the missing 
$S$-wave attraction without further sizable modifications, presumably because 
they change sign sharply very close to threshold (Fig.~\ref{fig:pwave}) which 
is the relevant energy in the present studies. 

We have also studied possible shifts, Eq.~(\ref{eq:SCmodifMS}), in the 
energy where the $P$-wave amplitudes are evaluated and found, in fact, 
that shifting to about 8~MeV lower energies greatly improves the agreement 
with experiment. However, judging by the $\bar pp$ and $\bar np$ annihilation 
cross sections, Figs.~\ref{fig:nbarann} and \ref{fig:pbarann} respectively, 
the energy scale cannot be changed by more than $\pm$1 MeV. 

Finally, the large discrepancies between measured and calculated 
$\bar n$-nucleus annihilation cross sections, observed earlier with 
empirical potentials \cite{Fri14,Fri15}, persist in the present 
work based on a more microscopic approach.

\section*{Acknowledgements}

We thank J.~Mare\v{s} and J. Hrt\'{a}nkov\'{a} for useful discussions.
SW was supported by NCN grant 2011/03/B/ST2/00270.

\section*{Appendix A.~~From repulsive free-space $\bar p N$ amplitudes to 
attractive in-medium $\bar p N$ amplitudes} 
\label{sec:appendix} 
\renewcommand{\theequation}{A.\arabic{equation}}
\setcounter{equation}{0}
\renewcommand{\thefigure}{A.\arabic{figure}} 
\setcounter{figure}{0} 

Here we demonstrate the WRW mechanism, Eq.~(\ref{eq:WRW}), responsible for 
turning repulsive free-space $\bar p N$ $S$-wave amplitudes into attractive, 
density dependent, in-medium $\bar p N$ amplitudes. It is instructive to 
estimate the size of the free-space $\bar p N$ repulsive $S$-wave cm 
amplitudes $f$ required to make the corresponding in-medium amplitudes 
$f_{\rm medium}(\rho)$ attractive somewhere at the nuclear surface and inward, 
say beginning at density $\rho=\rho_0/8$. For simplicity, we assume that 
$f_{\bar pp}=f_{\bar pn}$, so that Eq.~(\ref{eq:WRW}) at this particular value 
of density assumes the form 
\begin{equation} 
f_{\rm medium}(\rho=\rho_0/8)=\frac{f}{1+0.65f}, 
\label{eq:medium} 
\end{equation} 
where $f=f_R+if_I$ is given in fm. If $f$ is purely real ($f_I=0$) 
and repulsive ($f_R <0$), then $f_{\rm medium}(\rho > \rho_0/8)>0$, 
corresponding to in-medium attraction, for a sufficiently strong repulsive 
free-space amplitude with values of $f_R < -1.54$~fm. If $f$ is purely 
absorptive ($f_R=0$), then $f_{\rm medium}(\rho)$ is attractive 
and absorptive everywhere for any nonzero value of $f_I$. Taking 
a representative subthreshold value of $f_I=1$~fm from Fig.~\ref{fig:swave}, 
one gets $f_{\rm medium}(\rho=\rho_0/8)=(0.46+i 0.70)$~fm 
corresponding to a medium-size attraction and a somewhat reduced absorptivity. 
Increasing $f_I$, the in-medium attraction at $\rho_0/8$ increases up 
to a maximum value and the in-medium absorptivity continues to decrease. 
Finally, if $f$ is equally repulsive and absorptive ($-f_R=f_I$), 
Re\,$f_{\rm medium}(\rho > \rho_0/8)>0$ for values of $f_R < -0.77$~fm. 
Taking again a representative value of $f_I=1$~fm, one gets 
$f_{\rm medium}(\rho=\rho_0/8)=(0.55+i1.83)$~fm, again corresponding to 
a medium-size attraction, but to a sizable absorptivity enhanced by almost 
a factor of 2. Note from Fig.~\ref{fig:swave} that in order to satisfy the 
minimum-repulsion requirement of $f_R < -0.77$~fm, one has to go slightly 
below threshold.


\end{document}